\documentclass{aa}  

\usepackage{graphicx}
\usepackage{txfonts}
\begin{document}
	
	\title{Discovery of new stellar groups in the Orion complex}
	\subtitle{Towards a robust unsupervised approach}
	
	\author{Boquan Chen \inst{1,2,3}
          \and Elena D'Onghia \inst{2,4}
          \and Jo\~{a}o Alves \inst{5,6,7} 
          \and Angela Adamo \inst{8}}
          
    \institute{Sydney Institute for Astronomy, The University of Sydney, School of Physics A28, Camperdown, NSW 2006, Australia\\
                \email{bche3093@uni.sydney.edu.au} \and
                Department of Astronomy, University of Wisconsin, 475 North Charter Street, Madison, WI 53706, USA \and
                ARC Centre of Excellence for All Sky Astrophysics in Three Dimensions (ASTRO-3D) \and
                Center for Computational Astrophysics, Flatiron Institute, 162 5th Avenue, 10010 New York \and
                Department of Astrophysics, University of Vienna, T\"{u}rkenschanzstrasse 17, 1180 Wien, Austria \and
                Radcliffe Institute for Advanced Study, Harvard University, 10 Garden Street Cambridge, MA 02138, USA \and
                University of Vienna, Faculty of Computer Science, Data Science @ Uni Vienna \and
                Department of Astronomy, Stockholm University, Oscar Klein Centre, AlbaNova University Centre, 106 91, Stockholm, Sweden
                }
	
	\date{Received XXXX; accepted XXXX}
	
	\abstract
	{We test the ability of two unsupervised machine learning algorithms, \textit{EnLink} and Shared Nearest Neighbour (SNN), to identify stellar groupings in the Orion star-forming complex as an application to the 5-dimensional astrometric data from \textit{Gaia} DR2. The algorithms represent two distinct approaches to limiting user bias when selecting parameter values and evaluating the relative weights among astrometric parameters. \textit{EnLink} adopts a locally adaptive distance metric and eliminates the need of parameter tuning through automation. The original SNN relies only on human input for parameter tuning so we modified SNN to run in two stages. We first ran the original SNN 7,000 times, each with a randomly generated sample according to within-source co-variance matrices provided in \textit{Gaia} DR2 and random parameter values within reasonable ranges. During the second stage, we modified SNN to identify the most repeating stellar groups from 25,798 we obtained in the first stage. We reveal 21 spatially- and kinematically-coherent groups in the Orion complex, 12 of which previously unknown. The groups show a wide distribution of distances extending as far as about 150 pc in front of the star-forming Orion molecular clouds, to about 50 pc beyond them where we find, unexpectedly, several groups. Our results expose to view the wealth of sub-structure in the OB association, within and beyond the classical Blaauw Orion OBI sub-groups. A full characterization of the new groups is of the essence as it offers the potential to unveil how star formation proceeds globally in large complexes such as Orion. The data and code that generated the groups in this work as well as the final table can be found at \protect\url{ https://github.com/BoquanErwinChen/GaiaDR2_Orion_Dissection}.}
	
	
	\maketitle
	
	\section{Introduction}
	
    Disentangling between different young populations of similar ages in nearby star-forming regions promises to allow an accurate reconstruction of the star formation process of local giant molecular clouds and provide new insight into how young stellar populations form and disperse to build the Galactic field. Traditionally, distinguishing nearby young stellar populations with ages younger than $\sim100$ Myr has been a difficult task mostly because of the large solid angle in the sky that needs to be covered, and, when only photometry is available, sample contamination can severely hamper the analysis. Recently, the \textit{Gaia} mission has begun providing massive amounts of all-sky and high-quality photometry and astrometry, dramatically improving this situation. Clustering techniques are naturally becoming mainstream statistical tools for astronomers trying to identify populations of stars, but reproducibility can be problematic.
	
	An obvious target to disentangle young populations leaving their natal gas is the Orion complex, the closest massive star-forming region to Earth (see \citealt{bally2008,Alves2012,Bouy2014-fw,Bouy2015-ce,Kubiak2017-ur,Zari2017,Kounkel2018,Kos2018,Zari2019}). It is tempting to explore the Orion complex in the full 6d phase space, but unfortunately only a fraction of stars in the Complex have radial velocities available. 
	Recently, \cite{Kounkel2018} computed the distance matrices for stars with 3d (APOGEE \citep{Majewski2017}), 5D (\textit{Gaia} DR2), and 6d (APOGEE-Gaia) information separately and normalized them to produce a joint distance matrix. With a hierarchical clustering algorithm, they classified the Orion Complex into five components, Orion A, B, C, D, and $\lambda$ Ori. Similarly, \cite{Kos2018} introduced a custom metric in 6d phase space which contains a factor that makes distances calculated with 5D and 6d information compatible. Adopting an iterative approach instead, they identified 5 clusters in the Ori OB 1a association, including one estimated to be 21 Myrs old. The approaches in \cite{Kounkel2018} and \cite{Kos2018} both involve inexplicit assumptions about the missing dimension, radial velocity.
	
	In this work, we choose to omit radial velocities, given their limited availability, and focus on the 5D phase space in \textit{Gaia} DR2. Instead of scaling all dimensions to the same length, we test two algorithms that weigh the astrometric parameters in distinct ways. \textit{EnLink} \citep{enlink} partitions the entire sample into uniform chunks and uses a locally adaptive metric which is automatically adjusted by the algorithm for each chunk. For the second algorithm, Shared Nearest Neighbour (SNN) \citep{ertoz03}, we modified the original distance metric by introducing two parameters that control the spread in 3d spatial locations and 2d kinematics respectively. We will explore the parameter space extensively in order to remove user bias in parameter selection and measure the stability of our final groupings across a range of parameter values. 
	
	In Section \ref{data}, we will briefly describe our sample selection. In Section \ref{methodology}, we will describe our methodology in detail, including the heuristics behind the \textit{EnLink} and SNN clustering algorithms and parameter tuning for SNN. In Section \ref{results}, we will examine the stellar groups we recovered with both algorithms. The data and code that produced the results presented in this work can be found at \url{ https://github.com/BoquanErwinChen/GaiaDR2_Orion_Dissection}.

	\section{Data}
	\label{data}
	
	The data we use in this work will be \textit{Gaia} Data Release 2, which provides precise positions in the sky ($\alpha$, $\delta$), parallaxes ($\varpi$), and proper motions ($\mu_\alpha$, $\mu_\delta$), for over a billion stars \citep{GaiaDR2}. We adopt similar selection criteria in Gaia 5D astrometric parameters to \cite{Kounkel2018} to restrict our sample to the Orion complex, specifically $75^\circ < \alpha < 90^\circ$, $-15^\circ < \delta < 15^\circ$, $2 < \varpi < 5$ mas, $-4 < \mu_\alpha < 4$ mas/yr, $-4 < \mu_\delta < 4$ mas/yr. Instead of imposing cuts on errors, we limit our sources to those with re-normalised unit weight error (RUWE) < 1.4 (see Gaia technical note GAIA-C3-TN-LU-LL-124-01).  We do not further restrict our sample with color-magnitude cuts. A total of 29,030 stars are left for classification. \textit{Classification} is defined as clustering algorithms assigning valid (non-noise) group labels to stars. Unclassified stars are thus treated as noise. We feed the original 5d astrometric data to \textit{EnLink} because it weighs data dimensions through a co-variance matrix. For SNN, we convert the right ascension ($\alpha$), declination ($\delta$), and parallax of every star into 3d rectangular coordinates ($x$, $y$, $z$) with Astropy \citep{astropy:2013, astropy:2018}. We use the Distance module in Astropy to convert parallax into distance, though every model essentially returns the inverse of parallax. SNN takes ($x$, $y$, $z$) and proper motions as separate groups of input so we have two free parameters, $n_{xyz}$ and $d_{pm}$.
	
	\section{Methodology}
	\label{methodology}
	
	Henceforth, we refer to all structures recovered by clustering algorithms in the Orion field as \textit{stellar groups} or \textit{groups} to avoid confusion. One of the main challenges associated with clustering in our 5D phase space is how to find the ideal balance among the degrees of variation in astrometric parameters. Members of a stellar group could have a large spread in ($\alpha$, $\delta$, $\varpi$) but little spread in proper motions or the opposite. The most common approaches would be to normalize astrometric parameters to the same range or by their measurement errors and feed the normalized data to clustering algorithms as vectors. However, these approaches do not always promise the optimal balance for all stellar populations. Therefore, we use two distinct clustering algorithms, \textit{EnLink} and SNN. \textit{EnLink} will automatically pick the balance suitable to the geometry of each partition of our data space. As for SNN, we adopt a Monte Carlo approach to reveal the most likely stellar groups by running SNN 7,000 times and cross-matching them. Each time, we redraw our sample according to the within-source co-variance matrices provided in \textit{Gaia} DR2 and pick random parameter values. We then identify the most repeated groups through a cross-matching procedure derived from the same SNN algorithm in order to recover stellar groups with varying spatial and kinematic configurations.
	
	\subsection{\textit{EnLink}}
	
	\textit{EnLink} is a density-based hierarchical clustering algorithm and uses a locally adaptive Mahalanobis metric. The Mahalanobis distance is defined as 
	\begin{equation}\label{eq:3}
	d_M(\vec{x}, \vec{y}) = \sqrt{|\Sigma|^{1/d}(\vec{x}-\vec{y})\Sigma^{-1}(\vec{x}, \vec{y})(\vec{x}-\vec{y})^T} 
	\end{equation}
	\noindent where $d$ is the dimensionality of our data set, $\vec{x}$ and $\vec{y}$ are vectors in our 5D phase space for two stars, and $\Sigma(\vec{x}, \vec{y})$ is the co-variance matrix of data in local volumes containing $\vec{x}$ and $\vec{y}$ respectively. In practice, $\Sigma(\vec{x}, \vec{y})$ is approximated with $1/2(\Sigma(\vec{x}) + \Sigma(\vec{y}))$. $\Sigma(\vec{x})$ and $\Sigma(\vec{y})$ are calculated separately through a partitioning scheme such that points are as uniformly distributed as possible in each partition. The balance among our five astrometric parameters is determined in each of these local partitions through the co-variance matrices $\Sigma$. This metric is particularly useful when our groups have different scatters in separate dimensions. \textit{EnLink} assigns every point to a cluster by default and calculates a density score for each member in case outliers need to be removed. Since \textit{EnLink} minimizes the need for parameter selection, we use the default values for all parameters.Specifically, the number of nearest neighbours for density estimation is set to 10, the significance threshold to 5 $\sigma$, and the minimum group size to 10. The significance level of a group is defined as the ratio between the highest and lowest density of points in it and can be viewed as a signal-to-noise ratio. The adaptive metric of \textit{EnLink} makes it perfect for exploratory analysis, while SNN requires more knowledge of the data set and more effort in parameter tuning. 
	
	\subsection{SNN}
	\subsubsection{Introduction to SNN}
	\label{intro_snn}
	
	The Shared Nearest Neighbour (SNN) clustering algorithm can be viewed as a modified version of DBSCAN, short for Density-Based Spatial Clustering of Applications with Noise \citep{Ester96}. SNN inherits the same mechanism as DBSCAN, but adopts the Jaccard distance metric of nearest neighbours in 5D space to make the density threshold more flexible. The Jaccard distance is defined as  
	\begin{equation}\label{eq:1}
	d_J(A, B) = 1- \frac{|S_A \cap S_B|}{|S_A \cup S_B|} 
	\end{equation}
	\noindent where $S_A$ and $S_B$ represent the sets of neighbours for two stars, A and B, and the absolute value signs represent the cardinal/size of a set. If two stars share identical neighbours, $|S_A \cap S_B|$ and $|S_A \cup S_B|$ would be identical and thus the distance between them would be at minimum 0. If $A$ and $B$ share no neighbour, the distance between them would be at maximum 1. The Jaccard distance allows us to reduce high dimensional data into simple 1d arrays of nearest neighbours through more sophisticated means. The neighbours in these arrays are not necessarily neighbours of individual stars but also members of a stellar group which we will exploit to identify repeating groups in Section \ref{crosscheck}.
	
	
	\subsubsection{Steps in SNN} 
	
	SNN in general follows three steps: 1) Retrieve the nearest neighbours; 2) Compute the Jaccard distance and create a distance matrix; 3) Perform DBSCAN clustering with the pre-computed distance matrix. 
	
	Our modified SNN differs from the generic version in Step 1 where we first retrieve the same number of nearest neighbours for every star in rectangular spatial coordinates, $(x, y, z)$, and then prune those with dissimilar proper motions in order to keep neighbours that share both similar spatial locations and proper motions, an approach used for chemodynamical tagging in \citep{Chen2018}. Two parameters are thus involved in the selection of the nearest neighbours, $n_{xyz}$ and $d_{pm}$. $n_{xyz}$ is the number of nearest neighbours for every star in the rectangular coordinate converted from RA, Dec, and parallax from \textit{Gaia} DR2. $d_{pm}$ is the maximum difference allowed in proper motion vectors between a star and its nearest neighbours in spatial coordinates. Indeed, there are many other clustering algorithms available that mitigate the fixed density threshold in DBSCAN, such as OPTICS and HDBSCAN. However, without an appropriate metric to measure the proximity between stars in our 5D phase space, clustering algorithms are unlikely to obtain clusters in all dimensions. 
	
	Step 2 takes the lists of nearest neighbours from step 1 and converts them into a distance matrix by computing the Jaccard distance between every pair of stars. 
	
	\begin{figure*}[tbp!]
		\centering
		\includegraphics[width=0.9\linewidth]{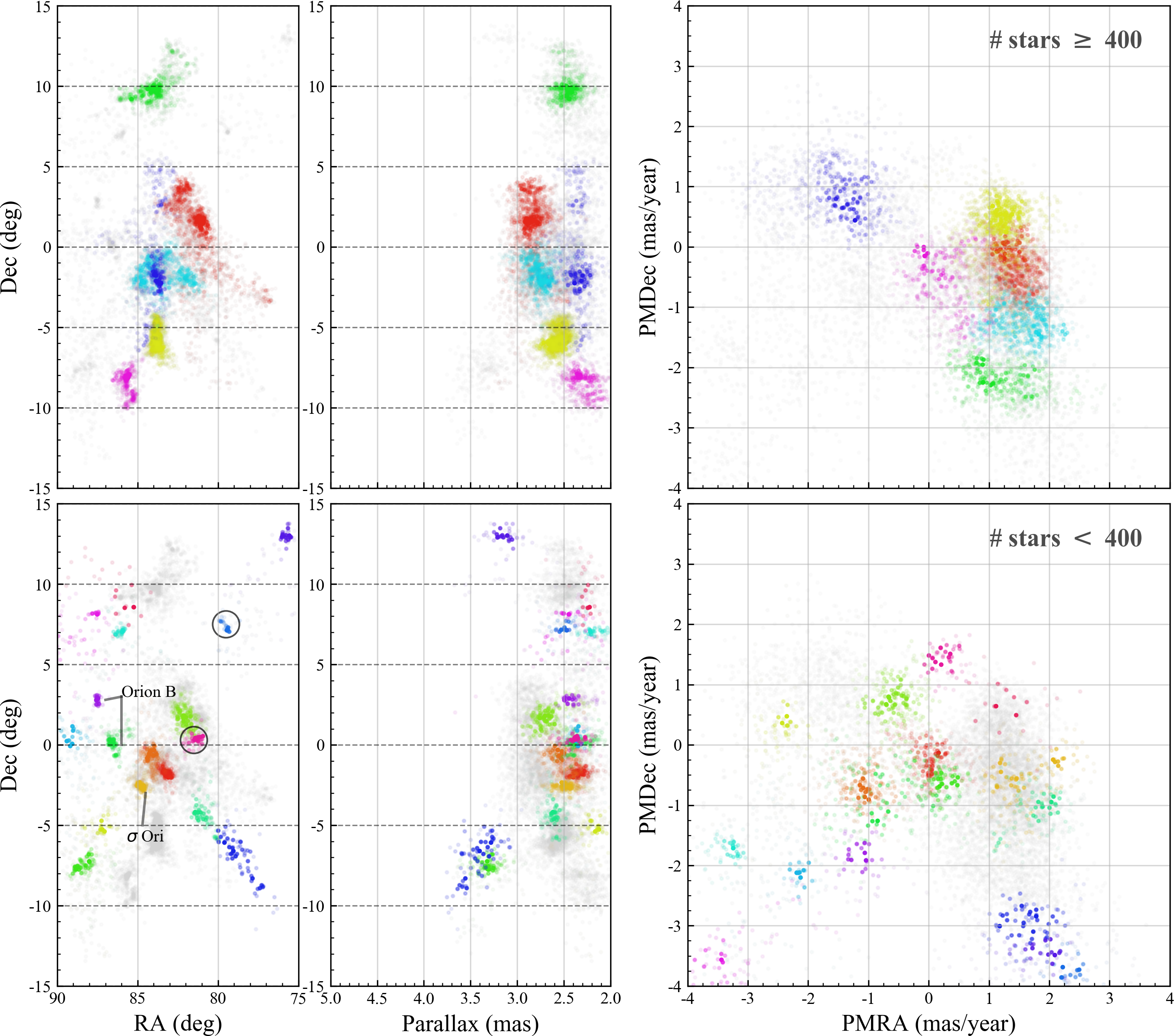}
		\caption{The 22 stellar groups recovered by \textit{EnLink}. The vertical columns from left to right show the groups in 2D projects of celestial coordinates (right ascension and declination), parallax and declination, and proper motions. Due to the high number of groups, we show groups with at least 400 stars in the first row and the rest in the second row. Within each row, each color represents a distinct stellar group recovered by \textit{EnLink}. The rest of the classified stars that do not fall into the size category in either row are shown as gray dots. \textit{EnLink} recovers several stellar groups not shown in SNN results, including $\sigma$ Ori, NGC 2024 and NGC 2068 as well as two spatially compact groups circled out in the sky in the second row. }
		\label{fig:orionradecenlink}
	\end{figure*}
	
	DBSCAN clustering comes in Step 3. Technically, any clustering algorithm that allows a pre-computed sparse distance matrix could be used in this step. However, we used DBSCAN because it was the original design of SNN \citep{ertoz03} and also because of its speed and proven ability to recover clusters of irregular shapes.
	
	\begin{figure*}[tbph!]
		\centering
		\includegraphics[width=0.9\linewidth]{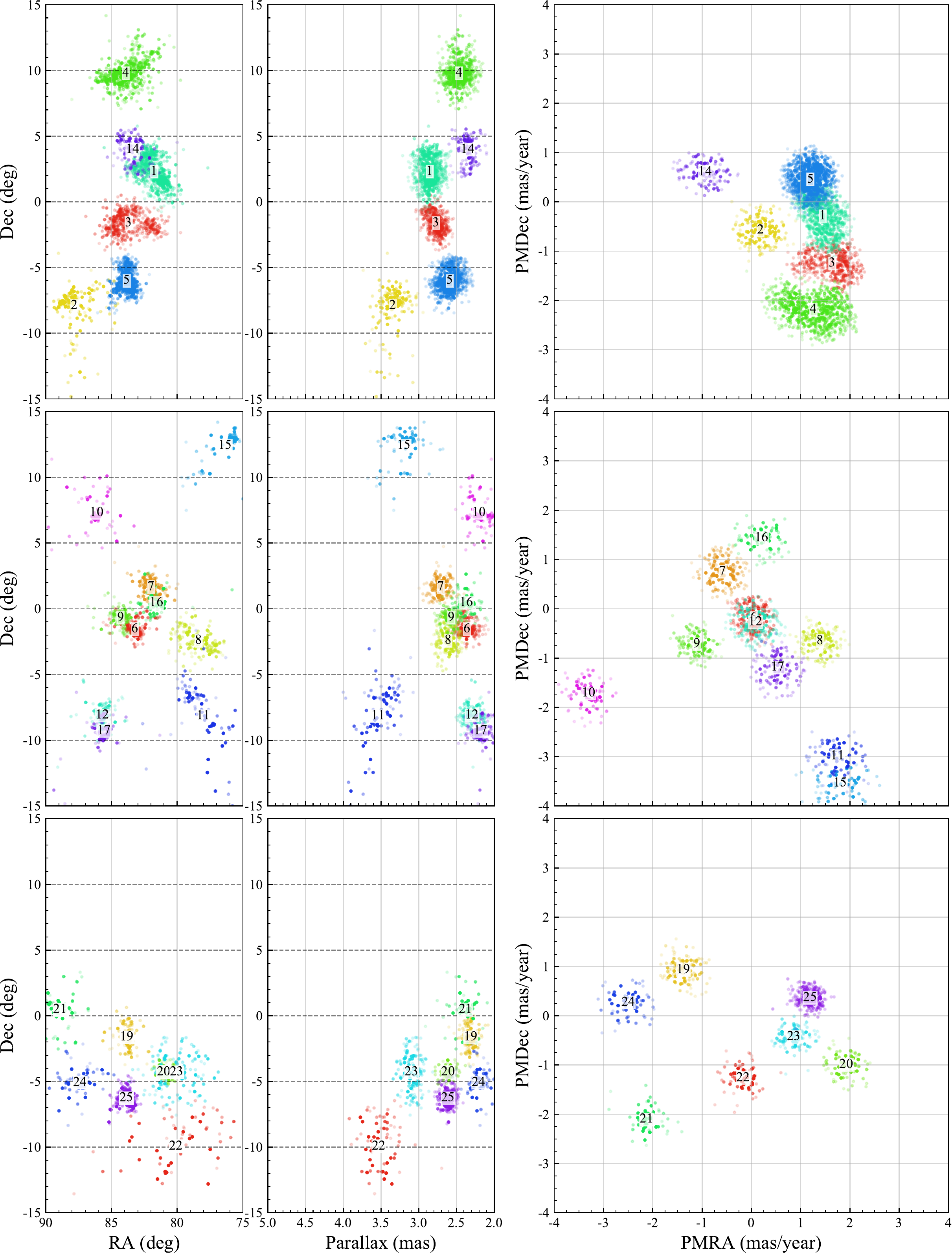}
		\caption{The 25 stellar groups recovered by SNN with our procedure to identify repeated groups in parameter space. Group 13 and 18 are omitted as they appear to be duplicates of Group 3. The organization of the panels is exactly the same as Fig. \ref{fig:orionradecenlink}. The groups are presented in three rows to provide the maximum separation in the three projections. Distinct colors in each row represent distinct stellar groups. Again, readers should note that the same color in different rows represents different groups. The stellar groups are assigned a numeric label in the last column, the proper motion projection. The same numeric labels are used in Table \ref{table:finaltable} to refer to each group. Compared to \textit{EnLink} results in Fig. \ref{fig:orionradecenlink}, the SNN stellar groups are more spatially spread out but have much tighter distributions in proper motion space. }
		\label{fig:orionradec}
	\end{figure*}
	
	\subsubsection{Parameter Tuning and Stellar Group Identification}
	\label{crosscheck}
	
    In this section, we will demonstrate how we simultaneously accomplish three tasks commonly involved in clustering analysis with a Monte Carlo approach: 1) parameter tuning, especially when one set of parameters is not sufficient to retrieve underlying structures; 2) calculate the frequentist stability of recovered groups, or the number of times a group appears; and 3) assign frequentist membership probability to group members, or the percentage of times a star appears in an assigned group. Our strategy is to run the SNN algorithm multiple times with randomly generated samples and parameter values and  then identify the most repeated groups as our final results. 
	
    We ran our SNN algorithm 7,000 times to find the underlying Orion stellar groups in our Orion sample. Each time, we redrew the astrometric parameters of the stars our sample according to the within-source 5$\times$ 5 co-variance matrices. We note that the number of SNN runs is set to 7,000 because we estimated the resulting Jaccard distance matrix would be of the maximum size that can be stored on 16 GB pf RAM. To fully explore the parameter space, we also redrew our algorithm parameter values from uniform distributions within the range of 50-1,400 and 0-0.6 for $n_{xyz}$ and $d_{pm}$ respectively. Given our sample size of 29,030 stars, the range adopted for $n_{x,y,z}$  allows every star a chance to connect to at most about 5\% of the sample closest to it in $(x, y, z)$. The maximum value of $d_{pm}$ is set to be twice the average error ($\approx$ 0.3 mas/year) of proper motions in either direction. 
    
    We fixed \textit{min\_samples} to be 20 and \textit{eps} to be 0.5, which mostly limits our classified stars to those with at least 20 stars with more than 50\% shared nearest neighbours in the 5D space with specified $n_{xyz}$ and $d_{pm}$. We chose the values for these two parameters through trial and error to minimize the number of groups resulting from noise in the data. When \textit{min\_samples} was lowered below 20, the number of stellar groups drastically increased to over one hundred. Many of these groups have very few stars ($\sim$ 20) and their spreads in proper motions are less than or comparable to measurement errors and thus are very likely caused by noise. As for \textit{eps}, we recover very few ($< 5$) stellar groups when its value is too low or too large.
    
    We have experimented with including \textit{eps} and/or \textit{min\_samples} as free parameters but it became substantially more computationally expensive to get a sufficient number of repeating groups for cross-matching later. Even after we increased the number of runs to 10,000, we obtained only a few thousand non-unique stellar groups, while we could easily get around 10,000 groups from just 2,000 runs if we chose to fix the values of \textit{min\_samples} and \textit{eps}. In practice, the values of $n_{xyz}$ and $d_{pm}$ also affect the density threshold set by \textit{min\_samples} and \textit{eps}. If we assume the members of a truly physical group follow a Gaussian distribution, we would retrieve increasingly less percentage of members as we extend search radius in 5d astrometric space by increasing $n_{xyz}$ and $d_{pm}$. Thus, we would need to lower \textit{eps} to eliminate the inclusion of unassociated stars and increase \textit{min\_samples} to accommodate the extra true members. We eventually decided to leave \textit{min\_samples} and \textit{eps} as fixed parameters.
    
     The 7,000 runs of SNN yielded 25,798 stellar groups which we cross-matched by utilising the same SNN algorithm. As mentioned before in Section \ref{intro_snn}, SNN translates the lists of neighbours of two objects into the Jaccard distance between them. These neighbours were previously stars in 5D astrometric space. However, if we substitute the lists of neighbouring stars with the lists of members of our 25,798 stellar groups, we transform SNN into a clustering algorithm on stellar groups. The meanings of \textit{eps} and \textit{min\_samples} are updated accordingly. \textit{eps} is now the minimum shared ratio of members between two groups and \textit{min\_samples} is the minimum number of groups with above such ratio. The values of \textit{eps} and \textit{min\_samples} are picked again through trial and error. In principle, we want to minimize the number of unclassified non-unique groups and avoid the same group from appearing multiple times due to some low-probability members. In the end, we picked \textit{eps}=0.7 and \textit{min\_samples}=15 to retrieve our final stellar groups. We could repeat this study with more computing resources when more precise data become available in the future. Thanks to parallel processes, the data analysis is completed on an 8th Gen Intel i7 8-core CPU in just one day.
    
     We define the frequentist stability as the number of times a group or its close duplicates appear out of 7,000 runs of SNN. During the second round of SNN, we selected the most stable 25 cross-matched stellar groups as our final SNN results, keeping the minimum stability above 10. We found that groups with a stability score less than 10 were sometimes replaced by other groups when we reran the analysis. We saved only members with at least 10\% probability to avoid confusion due to fortuitous overlaps, which left us with 4,598 unique stars. As mentioned before, a star could be assigned to multiple groups.
	
	\section{Results}
	\label{results}
	
	\subsection{Classified Stars}
	
	We will first present the results from \textit{EnLink} as the initial exploratory analysis. Since \textit{EnLink} assigns every star to a stellar group, it is necessary for us to remove outliers with very low density score. We removed stars with bottom 10\% density score overall and kept only the 30\% of the members with the highest density score in each group to reveal its core because of large overlaps between groups in proper motion space. 
	
	\begin{figure}[tbph!]
		\centering
		\includegraphics[width=0.95\linewidth]{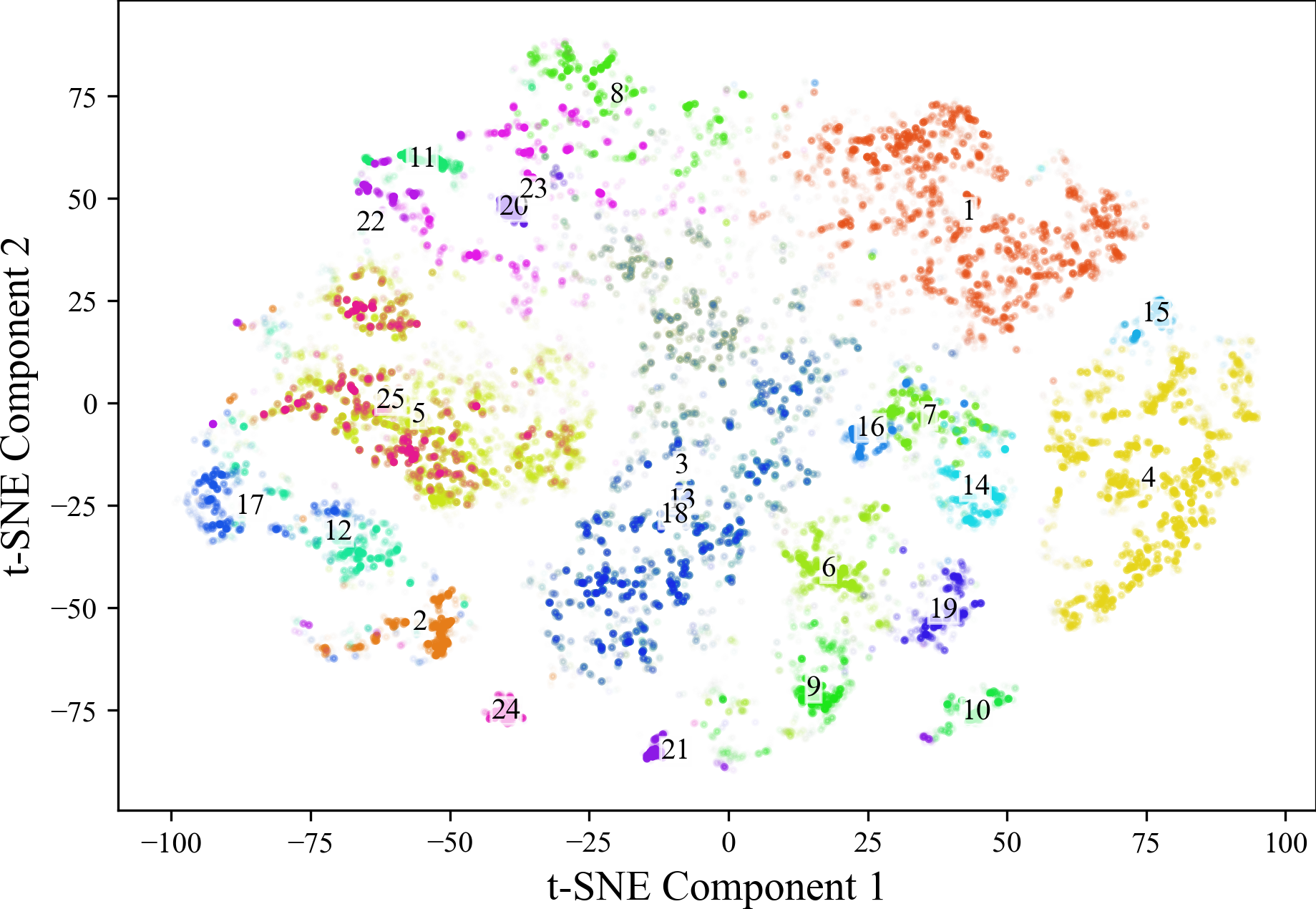}
		\caption{SNN stellar groups in t-SNE 2D projection. Each stellar group is highlighted with a distinct color and a numeric label. The groups are coherent in this 2D space reduced from 5D astrometric space, affirming our results obtained from SNN. }
		\label{fig:tsne}
	\end{figure}
	
	A total of 8,699 stars remains after the selection in density score. Figure \ref{fig:orionradecenlink} shows all 22 \textit{EnLink} stellar groups in three 2D projections: RA and Dec, parallax and Dec, and proper motions. The groups are divided into two categories: those with at least 400 stars are shown in the first row and the rest are shown in the second row. Readers should note that we reuse colors across the rows to maximize the separation among stellar groups in color space. The opacity of the dots are coded by density scores returned by \textit{EnLink}. Even after our cut, many stars still have very low density scores. The large groups in the first row mostly agree with the classification in \cite{Kounkel2018}, though here they contain much less stars. 
	
	The small groups in the second row reveal that the automatic adaptive metric in \textit{EnLink} assigns much more weight to RA, Dec and parallax than proper motions. The groups are much more coherent in the first two projections than in proper motion space. Because of this, \textit{EnLink} recovers NGC 2024 (green) and NGC 2068 (purple) in Orion B and $\sigma$ Ori (Orange) as well as two other spatially compact groups which are missing in SNN results. SNN judged these groups unstable because of their large spreads in proper motion space. We are not overly concerned with SNN not recovering NGC 2024 and NGC 2068 because the goal of this work is to find new stellar groups. As for the two spatially compact groups, SNN recovered dozens of unstable groups similar to these so they are quite common occurrences with SNN. We decided not to discuss them in this work and focus on the most stable stellar groups.
	
	\begin{figure*}[h!]
		\centering
		\includegraphics[width=0.9\linewidth]{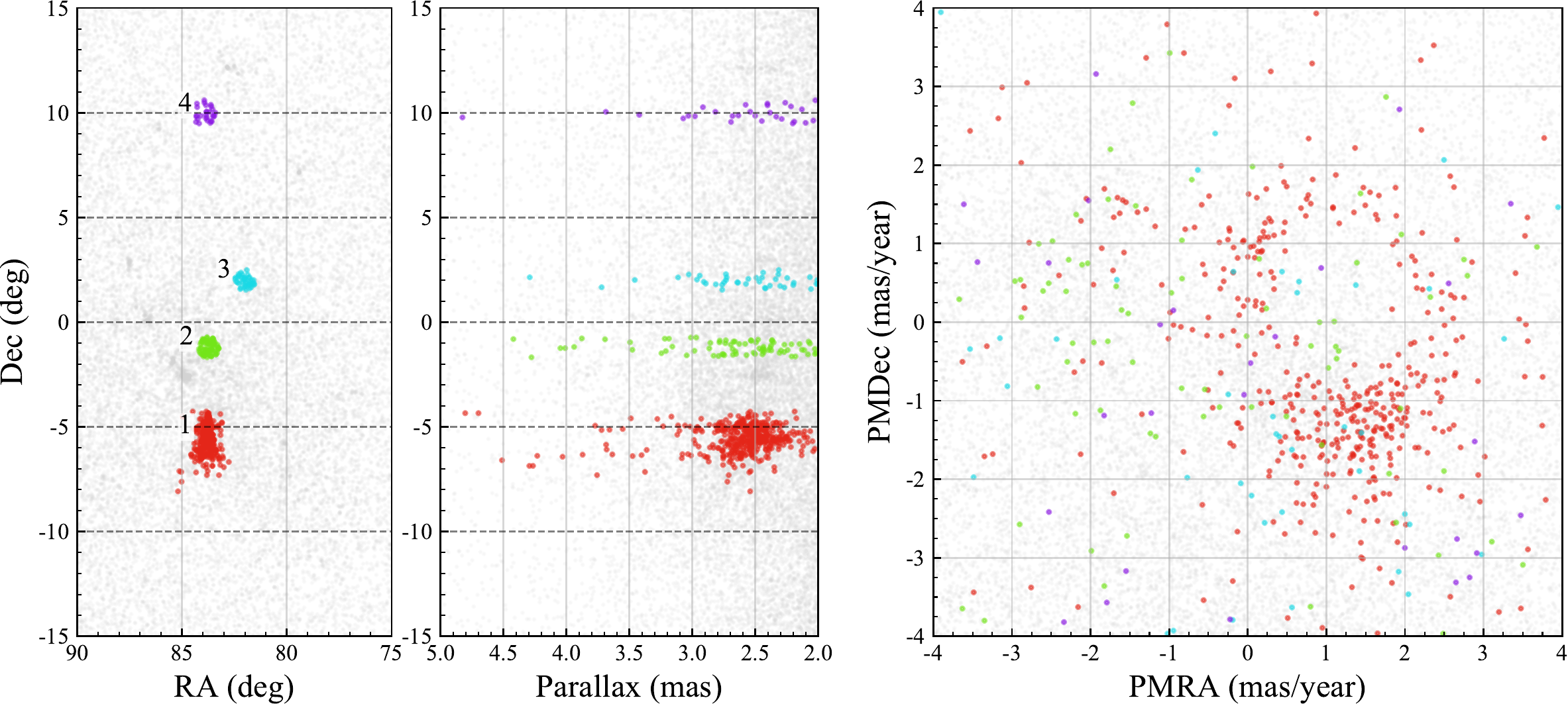}
		\caption{Stars unclassified by SNN clustering algorithm. Stars at four over-dense regions in the sky are highlighted in distinct colors and examined across the same projections as Fig. \ref{fig:orionradecenlink}. Spot 1 (red) targets the head of Orion A where Group 5 is recovered. A hole is visible in the proper motion space due to the classified stars in Group 5. Spot 2 and 3 targets the OB1b and OB1a associations. Spot 4 targets $\lambda$ Ori. No over-density can be identified in these spots. We omit over-densities that correspond to $\sigma$ Ori and Orion B here because they are already identified in Fig. \ref{fig:orionradecenlink}. }
		\label{fig:orionradecbg}
	\end{figure*}
	
	Now we will discuss the stellar groups from our two-stage SNN analysis as our final results. Figure \ref{fig:orionradec} shows the SNN stellar groups in the same projections as Fig. \ref{fig:orionradecenlink}. This time we place the groups into three rows to maximize their separation in 5D space instead of separating them by size. Again we need to reuse colors to make these stellar groups distinguishable by eye. The opacity of points in Fig. \ref{fig:orionradec} is coded by its frequentist membership probability and each group is assigned a numeric label according to its order of frequentist stability (Group 1 is the most stable) for easier identification. SNN was able to recover almost all \textit{EnLink} stellar groups, albeit the SNN counterparts appear more coherent in proper motion space. In Figure.~\ref{fig:tsne}, we further confirm that our SNN groups are conherent in 5D astrometric space by mapping the groups onto a 2D space with t-distributed Stochastic Neighbour Embedding (t-SNE) \citep{tsne}. t-SNE is a common dimension reduction technique to help visualize structures in high-dimensional data and has been widely applied in astronomy (see \cite{Lin2019galah, Zhang2019tsne, Kos2018}). The most important feature of both SNN and \textit{EnLink} results is perhaps that most groups correspond to known structures in the Orion Complex, which gives us confidence that the extraction of groups is physically meaningful.
	
	An important implication of our results is finding that there are several distinct groups along the same line of sight, consistent to the argument in \cite{Alves2012}. This is revealing of the overall architecture of the region, but poses an obvious complications for studies of this region. An example is the area of the Orion Belt Population, or OBP \citep{Kubiak2017-ur}, where at least five groups can be seen along the same general line-of-sight towards the Orion Belt stars with distinct proper motions. Another example, Groups 6, 9, and 16 in the second row of Fig.~\ref{fig:orionradec} are adjacent in spatial coordinates but do not overlap at all in proper motion space. Comparing to the results in \cite{Kounkel2018}, we find that Orion D in is divided at least among Groups 1, 7, 8, 9, 22, and 23 in our results. These groups are distinct enough in proper motions space to be identified as individual groups, revealing a complex dynamical structure. Some of these groups were already identified in the past: Group 1 clearly clusters around 25 Ori \citep{briceno07} and Group 23 matches Orion X \citep{Bouy2015-ce}. Follow-up work will investigate what these differences might mean in a global analysis of kinematics in the region. 
	
	We will now discuss some of the more noteworthy groups:
	
	\begin{itemize}
	
	    \item Group 1 coincides with the 25 Ori cluster \citep{briceno07,Briceno2018}, although it appears much more elongated than previously assumed, reaching as far South as L1616. We note, however, that the distance to this group is about 350 pc, i.e., about 100 pc further away than the  distance to the 25 Ori star, meaning that the star is probably is not part of the cluster. To avoid confusion in the future, where better data will become available, we suggest naming the group as Briceno 1, following the discovery paper \cite{briceno07}.
	
	    \item Group 3 appears to be one of the richer groups as it contains Group 13 and 18. However, we do not observe any significant offset among the three groups in the five input dimensions and thus choose to omit Group 13 and 18. At a distance of about 360 pc, Group 3 lies towards the Orion Belt Population \citep{Kubiak2017-ur} but is located towards the front, hence the OBP-near. 
	    
	    
	   \item Group 4 corresponds to the well known $\lambda$ Ori cluster but  new group, Group 10, which we name $\lambda$ Ori South, is clearly identified in the south of $\lambda$ Ori and lies slightly beyond. 
	    
	    \item Group 5, or the NGC1980 group, is one of the richer groups and it contains the disputed foreground population to the ONC. The group peaks around NGC 1980 and was discovered in \cite{Alves2012} and further extended in \citep{Bouy2014-fw} (see \cite{Da_Rio2016-iw,Fang2017-rr,Kounkel2017-rh,Beccari2017-yn,Jerabkova2019-qo} for a different view). The group extends roughly from NGC1981 down to South of NGC1980, passing in front of the ONC, and in particular, the Trapezium cluster, which we do not manage to retrieve from Gaia DR2. The foreground to the ONC and the Trapezium is an important one, as this cluster is a benchmark for star formation studies, and an unrecognized foreground population will compromise the basic properties derived for the region. Better data in Gaia EDR3, and in particular, the large number of radial velocities available in Gaia DR3, should be able to clarify this situation and the actual position of Group 5 in relation to the ONC. 
	    
	    \item Group 7 corresponds to a layer in Blaauw's subgroup Ia and was identified as ASCC 20 by \cite{Kos2018}. The coordinates and parallax values of Group 7 also resemble the young population found in \cite{Zari2017}. 
	    
	    \item Group 8 lies below the Orion Belt Population and extends to L1616, a small cometary cloud at the same distance and currently star-forming (at least one source in the sample appears to belong to the small star-forming region). It is possible that this group formed from a cloud now gone except for the leftover cometary cloud L1616, but more data would be needed to make this conclusion.  
	    
	    \item Group 11 seems to trace IC 2118, a faint and elongated reflection nebula of Rigel. It is likely that Rigel is part of this group as it overlaps with it in projection and is at about the same distance within the Hipparcos parallax error (the reason we name it here as the Rigel Group), but this statement needs confirmation in future Gaia data releases. Group 22 is classified as a subgroup of this group and captures the low density members. 
	    
	   \item Group 12, or the L1641S group, is associated with ongoing star formation towards the tail of the cloud, in particular, the L1641S cluster, as well as two smaller clusters towards the Southern tail of the cloud (L1647, see \cite{Strom93,Meingast2016-nb,Grossschedl2019}).
	   
	   \item Group 14, or the ome Ori group, is a dispersed group at about 417 pc, between Orion B and the $\lambda$ Ori molecular ring. ome Ori, a Be star with a reflection nebula is part of the group and it at about the average distnace tot he Group. 
	   
	   \item Group 15, or L1562, to the West of the $\lambda$ Ori group, lies at a distance of about 317 pc, i.e., about 100 pc in front of $\lambda$ Ori, and might be associated with some of the leftover molecular material in the region, such as the small cometary clouds L1571 and the  smaller one , L1562 at $l,b=(187.2,-16.7)$, which lies close to the main clustering of sources in this group.
	   
	   \item Group 16, or OBP-West, is a new, well defined and centrally concentrated group at a distance of about 416 pc. It overlaps extensively with Group 1 (25 Ori) but lies about 60 pc towards the back of the 25 Ori group. It is located West of the Orion Belt Population, and overlaps slightly with it, so we name it OBP-West.
	   
	   \item Group 17 seems to coincide with the partially embedded young populations at the tail of Orion A, L1647, hence the name of the group, at a distance of about 450 pc \citep{Grosschedl2018-mm,Grossschedl18-3d}. 
	   
	   \item Group 19, OBP-far, is the furthest group along the line of sight to the Orion Belt, at a distance of about 426 pc. 
	   
	   \item Group 20 includes the stellar yield of the L1634 cloud (for example, \cite{Alcala2008-va}), but includes a previously unknown population 1.5 degree northwest of L1634, hence our naming, the L1634 group. 
	   
	   \item Group 21 (Orion B West) and 24 (Orion A West) are dispersed groups to the West of Orion A and B clouds and are among the groups with lower stability. Group 24 is the most distance of all groups in the region, surprisingly, beyond the Orion A cloud. If this cluster is young and formed in the larger Orion star forming region, it will constitute an important clue that star formation in Orion did not happen in a linear and coherent manner, one producing a well defined age gradient, but it will likely have happen more chaotically.  
	    
	\end{itemize}
    
	\subsection{Unclassified Stars}
    \label{unclassified}
    
    Since SNN is based on DBSCAN which leaves out data points as noise, we will briefly discuss the unclassified stars. Figure \ref{fig:orionradecbg} shows the unclassified stars in the same projection as Fig. \ref{fig:orionradecenlink} and \ref{fig:orionradec}. All unclassified stars are shown in gray by default, while those at four over-dense regions in the sky are highlighted in distinct colors. We want to understand why SNN did not assign labels to these stars and possibly identify any groups that SNN missed, besides the ones already found by \textit{EnLink}.
    
   Spot 1 targets the head of Orion A where Group 5 is recovered. A large number of stars were not picked up by SNN in this region. A hole is visible in the proper motion distribution due to the classified stars in Group 5. Therefore, SNN treated these stars as noise primarily because they are outliers in proper motions. Spot 2 and 3 target the OB1b and OB1a associations. Spot 4 targets the leftover stars from $\lambda$ Ori. Similar to the stars in Spot 1, no over-density can be visually identified in proper motion space from these spots, even if we plot the spots individually to remove obscuring effect due to overlap. Overall, SNN did a satisfactory job removing over-densities from the astrometric 5D space.

	\section{Conclusion}
	\label{conclusion}
	
	We used the parameter-free \textit{EnLink} as an exploratory tool and adopted SNN algorithm to dissect the Orion star-forming complex. Our two-stage SNN procedure incorporates the within-source co-variance matrices and accomplishes three tasks commonly involved in clustering analysis, i.e. parameter tuning, computing group stability and assigning frequentist membership probability. We recovered 21 spatially- and kinematically-coherent unique stable groups in the Orion complex, as Group 11, 13, 18, and 25 appear to be only the highest or lowest density regions of their parent groups. Perhaps more remarkable, most of the unique groups identified in this work match previously well-known stellar populations, which gives us confidence in the approach. We found 12 new stellar groups, spread as far as about 150 pc in front of the star-forming Orion clouds, to about 50 pc beyond them, where we find, unexpectedly several groups, revealing the wealth of sub-structure in the OB association, within and beyond the classical Blaauw Orion OBI sub-groups \cite{Blaauw1964-tz,Brown1994-nu,De_Zeeuw1999-az}.

    
    The analysis in this work should be repeated including the sixth dimension, radial velocity, in the future. We expect radial velocity to be available for more stars in the Orion complex from Gaia DR3 or ongoing GALAH and APOGEE observations. The addition of radial velocity will allow us to produce stellar groups consistent in 6d phase space and study the kinematics of these groups more confidently. Gaia DR3 will provide more precise parallaxes and proper motions, which would further improve clustering results and allow us to identify finer structures in Orion. For now, a full characterization of the new groups is of the essence as it offers the potential to unveil how star formation proceeds globally in large complexes such as Orion. 
    
    \section{Acknowledgements}
	\label{acknowledgement}
    We thank the anonymous referee and Karolina Kubiak for their useful comments. B.C. is supported by the Research Training Program (RTP) offered by the Australian Department of Education. A.A. acknowledges the support of the Swedish Research Council, Vetenskapsr{\aa}det, and the Swedish National Space Agency (SNSA). This work has made use of data from the European Space Agency (ESA) mission Gaia (https://www.cosmos.esa.int/Gaia), processed by the Gaia Data Processing and Analysis Consortium (DPAC, https://www.cosmos.esa.int/web/Gaia/ dpac/consortium). Funding for the DPAC has been provided by national institutions, in particular the institutions participating in the Gaia Multilateral Agreement. This project was developed in part at the 2018 NYC Gaia Sprint, hosted by the Center for Computational Astrophysics of the Flatiron Institute in New York City and in part at the 2019 Santa Barbara Gaia Sprint, hosted by the Kavli Institute for Theoretical Physics at the University of California, Santa Barbara. This research was supported in part at KITP by the Heising-Simons Foundation and the National Science Foundation under Grant No. NSF PHY-1748958.

	\bibliographystyle{aa}
	\bibliography{ref}

\begin{thebibliography}{38}
\expandafter\ifx\csname natexlab\endcsname\relax\def\natexlab#1{#1}\fi

\bibitem[{Alcal{\'a} {et~al.}(2008)Alcal{\'a}, Covino, \&
  Leccia}]{Alcala2008-va}
Alcal{\'a}, J.~M., Covino, E., \& Leccia, S. 2008, Handbook of Star Forming
  Regions, 64, 801

\bibitem[{{Alves} \& {Bouy}(2012)}]{Alves2012}
{Alves}, J. \& {Bouy}, H. 2012, \aap, 547, A97

\bibitem[{{Astropy Collaboration} {et~al.}(2013){Astropy Collaboration},
  {Robitaille}, {Tollerud}, {Greenfield}, {Droettboom}, {Bray}, {Aldcroft},
  {Davis}, {Ginsburg}, {Price-Whelan}, {Kerzendorf}, {Conley}, {Crighton},
  {Barbary}, {Muna}, {Ferguson}, {Grollier}, {Parikh}, {Nair}, {Unther},
  {Deil}, {Woillez}, {Conseil}, {Kramer}, {Turner}, {Singer}, {Fox}, {Weaver},
  {Zabalza}, {Edwards}, {Azalee Bostroem}, {Burke}, {Casey}, {Crawford},
  {Dencheva}, {Ely}, {Jenness}, {Labrie}, {Lim}, {Pierfederici}, {Pontzen},
  {Ptak}, {Refsdal}, {Servillat}, \& {Streicher}}]{astropy:2013}
{Astropy Collaboration}, {Robitaille}, T.~P., {Tollerud}, E.~J., {et~al.} 2013,
  \aap, 558, A33

\bibitem[{{Bally}(2008)}]{bally2008}
{Bally}, J. 2008, {Overview of the Orion Complex}, ed. B.~{Reipurth}, Vol.~4,
  459

\bibitem[{Beccari {et~al.}(2017)Beccari, Petr-Gotzens, Boffin, Romaniello,
  Fedele, Carraro, De~Marchi, de~Wit, Drew, Kalari, Manara, Martin, Mieske,
  Panagia, Testi, Vink, Walsh, \& Wright}]{Beccari2017-yn}
Beccari, G., Petr-Gotzens, M.~G., Boffin, H. M.~J., {et~al.} 2017, Astron.
  Astrophys. Suppl. Ser., 604, A22

\bibitem[{Blaauw(1964)}]{Blaauw1964-tz}
Blaauw, A. 1964, Annu. Rev. Astron. Astrophys., 2, 213

\bibitem[{Bouy {et~al.}(2014)Bouy, Alves, Bertin, Sarro, \&
  Barrado}]{Bouy2014-fw}
Bouy, H., Alves, J., Bertin, E., Sarro, L.~M., \& Barrado, D. 2014, Astron.
  Astrophys. Suppl. Ser., 564, A29

\bibitem[{Bouy \& Alves(2015)}]{Bouy2015-ce}
Bouy, H. \& Alves, J.~F. 2015, Astron. Astrophys. Suppl. Ser., 584, A26

\bibitem[{{Brice{\~n}o} {et~al.}(2007){Brice{\~n}o}, {Hartmann},
  {Hern{\'a}ndez}, {Calvet}, {Vivas}, {Furesz}, \& {Szentgyorgyi}}]{briceno07}
{Brice{\~n}o}, C., {Hartmann}, L., {Hern{\'a}ndez}, J., {et~al.} 2007, \apj,
  661, 1119

\bibitem[{{Briceno} {et~al.}(2018){Briceno}, {Calvet}, {Hernandez}, {Vivas},
  {Mateu}, {Downes}, {Loerincs}, {Perez-Blanco}, {Berlind}, {Espaillat},
  {Allen}, {Hartmann}, {Mateo}, \& {Bailey}}]{Briceno2018}
{Briceno}, C., {Calvet}, N., {Hernandez}, J., {et~al.} 2018, arXiv e-prints,
  arXiv:1805.01008

\bibitem[{{Brown} {et~al.}(1994){Brown}, {de Geus}, \& {de
  Zeeuw}}]{Brown1994-nu}
{Brown}, A.~G.~A., {de Geus}, E.~J., \& {de Zeeuw}, P.~T. 1994, \aap, 289, 101

\bibitem[{{Chen} {et~al.}(2018){Chen}, {D'Onghia}, {Pardy}, {Pasquali},
  {Bertelli Motta}, {Hanlon}, \& {Grebel}}]{Chen2018}
{Chen}, B., {D'Onghia}, E., {Pardy}, S.~A., {et~al.} 2018, \apj, 860, 70

\bibitem[{Da~Rio {et~al.}(2016)Da~Rio, Tan, Covey, Cottaar, Foster, Cullen,
  Tobin, Kim, Meyer, Nidever, Stassun, Drew~Chojnowski, Flaherty, Majewski,
  Skrutskie, Zasowski, \& Pan}]{Da_Rio2016-iw}
Da~Rio, N., Tan, J.~C., Covey, K.~R., {et~al.} 2016, ApJ, 818, 59

\bibitem[{de~Zeeuw {et~al.}(1999)de~Zeeuw, Hoogerwerf, de~Bruijne, Brown, \&
  Blaauw}]{De_Zeeuw1999-az}
de~Zeeuw, P.~T., Hoogerwerf, R., de~Bruijne, J. H.~J., Brown, A. G.~A., \&
  Blaauw, A. 1999, Astron. J., 117, 354

\bibitem[{Ertoz {et~al.}(2003)Ertoz, Steinbach, \& Kumar}]{ertoz03}
Ertoz, L., Steinbach, M., \& Kumar, V. 2003, in Proceedings in Applied
  Mathematics, Vol. 112, Proceedings of the Third SIAM International Conference
  on Data Mining (SDM 2003), ed. D.~Barbara \& C.~Kamath (Society for
  Industrial and Applied Mathematics)

\bibitem[{Ester {et~al.}(1996)Ester, Kriegel, Sander, \& Xu}]{Ester96}
Ester, M., Kriegel, H.-P., Sander, J., \& Xu, X. 1996, in Proceedings of the
  Second International Conference on Knowledge Discovery and Data Mining,
  KDD'96 (AAAI Press), 226--231

\bibitem[{Fang {et~al.}(2017)Fang, Kim, Pascucci, Apai, Zhang, Sicilia-Aguilar,
  Alonso-Mart{\'\i}nez, Eiroa, \& Wang}]{Fang2017-rr}
Fang, M., Kim, J.~S., Pascucci, I., {et~al.} 2017, AJS, 153, 188

\bibitem[{{Gaia Collaboration} {et~al.}(2018){Gaia Collaboration}, {Brown},
  {Vallenari}, {Prusti}, {de Bruijne}, {Babusiaux}, {Bailer-Jones}, {Biermann},
  {Evans}, {Eyer}, \& et~al.}]{GaiaDR2}
{Gaia Collaboration}, {Brown}, A.~G.~A., {Vallenari}, A., {et~al.} 2018, \aap,
  616, A1

\bibitem[{Gro{\ss}schedl {et~al.}(2018)Gro{\ss}schedl, Alves, Meingast, Ackerl,
  Ascenso, Bouy, Burkert, Forbrich, F{\"u}rnkranz, Goodman, Hacar, Herbst-Kiss,
  Lada, Larreina, Leschinski, Lombardi, Moitinho, Mortimer, \&
  Zari}]{Grosschedl2018-mm}
Gro{\ss}schedl, J.~E., Alves, J., Meingast, S., {et~al.} 2018, Astron.
  Astrophys. Suppl. Ser., 619, A106

\bibitem[{{Gro{\ss}schedl} {et~al.}(2018){Gro{\ss}schedl}, {Alves}, {Meingast},
  \& {Hasenberger}}]{Grossschedl18-3d}
{Gro{\ss}schedl}, J.~E., {Alves}, J., {Meingast}, S., \& {Hasenberger}, B.
  2018, arXiv e-prints, arXiv:1812.08024

\bibitem[{{Gro{\ss}schedl} {et~al.}(2019){Gro{\ss}schedl}, {Alves}, {Teixeira},
  {Bouy}, {Forbrich}, {Lada}, {Meingast}, {Hacar}, {Ascenso}, {Ackerl},
  {Hasenberger}, {K{\"o}hler}, {Kubiak}, {Larreina}, {Linhardt}, {Lombardi}, \&
  {M{\"o}ller}}]{Grossschedl2019}
{Gro{\ss}schedl}, J.~E., {Alves}, J., {Teixeira}, P.~S., {et~al.} 2019, \aap,
  622, A149

\bibitem[{Jerabkova {et~al.}(2019)Jerabkova, Beccari, Boffin, Petr-Gotzens,
  Manara, Moroni, Tognelli, \& Degl'Innocenti}]{Jerabkova2019-qo}
Jerabkova, T., Beccari, G., Boffin, H. M.~J., {et~al.} 2019
  [\eprint[arXiv]{1905.06974}]

\bibitem[{{Kharchenko} {et~al.}(2013){Kharchenko}, {Piskunov}, {Schilbach},
  {R{\"o}ser}, \& {Scholz}}]{Kharchenko13}
{Kharchenko}, N.~V., {Piskunov}, A.~E., {Schilbach}, E., {R{\"o}ser}, S., \&
  {Scholz}, R.~D. 2013, \aap, 558, A53

\bibitem[{{Kos} {et~al.}(2018){Kos}, {Bland-Hawthorn}, {Freeman}, {Buder},
  {Traven}, {De Silva}, {Sharma}, {Asplund}, {Duong}, {Lin}, {Lind}, {Martell},
  {Simpson}, {Stello}, {Zucker}, {Zwitter}, {Anguiano}, {Da Costa}, {D'Orazi},
  {Horner}, {Kafle}, {Lewis}, {Munari}, {Nataf}, {Ness}, {Reid}, {Schlesinger},
  {Ting}, \& {Wyse}}]{Kos2018}
{Kos}, J., {Bland-Hawthorn}, J., {Freeman}, K., {et~al.} 2018, \mnras, 473,
  4612

\bibitem[{{Kounkel} {et~al.}(2018){Kounkel}, {Covey}, {Su{\'a}rez},
  {Rom{\'a}n-Z{\'u}{\~n}iga}, {Hernandez}, {Stassun}, {Jaehnig}, {Feigelson},
  {Pe{\~n}a Ram{\'\i}rez}, {Roman-Lopes}, {Da Rio}, {Stringfellow}, {Kim},
  {Borissova}, {Fern{\'a}ndez-Trincado}, {Burgasser},
  {Garc{\'\i}a-Hern{\'a}ndez}, {Zamora}, {Pan}, \& {Nitschelm}}]{Kounkel2018}
{Kounkel}, M., {Covey}, K., {Su{\'a}rez}, G., {et~al.} 2018, \aj, 156, 84

\bibitem[{Kounkel {et~al.}(2017)Kounkel, Hartmann, Calvet, \&
  Megeath}]{Kounkel2017-rh}
Kounkel, M., Hartmann, L., Calvet, N., \& Megeath, T. 2017, AJS, 154, 29

\bibitem[{Kubiak {et~al.}(2017)Kubiak, Alves, Bouy, Sarro, Ascenso, Burkert,
  Forbrich, Gro{\ss}schedl, Hacar, Hasenberger, Lombardi, Meingast, K{\"o}hler,
  \& Teixeira}]{Kubiak2017-ur}
Kubiak, K., Alves, J., Bouy, H., {et~al.} 2017, Astron. Astrophys. Suppl. Ser.,
  598, A124

\bibitem[{{Lin} {et~al.}(2019){Lin}, {Asplund}, {Ting}, {Casagrand e}, {Buder},
  {Bland-Hawthorn}, {Casey}, {De Silva}, {D'Orazi}, {Freeman}, {Kos}, {Lind},
  {Martell}, {Sharma}, {Simpson}, {Zwitter}, {Zucker}, {Minchev},
  {{\v{C}}otar}, {Hayden}, {Horner}, {Lewis}, {Nordlander}, {Wyse}, \&
  {{\v{Z}}erjal}}]{Lin2019galah}
{Lin}, J., {Asplund}, M., {Ting}, Y.-S., {et~al.} 2019, \mnras, 2724

\bibitem[{{Majewski} {et~al.}(2017){Majewski}, {Schiavon}, {Frinchaboy},
  {Allende Prieto}, {Barkhouser}, {Bizyaev}, {Blank}, {Brunner}, {Burton},
  {Carrera}, {Chojnowski}, {Cunha}, {Epstein}, {Fitzgerald}, {Garc{\'\i}a
  P{\'e}rez}, {Hearty}, {Henderson}, {Holtzman}, {Johnson}, {Lam}, {Lawler},
  {Maseman}, {M{\'e}sz{\'a}ros}, {Nelson}, {Nguyen}, {Nidever}, {Pinsonneault},
  {Shetrone}, {Smee}, {Smith}, {Stolberg}, {Skrutskie}, {Walker}, {Wilson},
  {Zasowski}, {Anders}, {Basu}, {Beland}, {Blanton}, {Bovy}, {Brownstein},
  {Carlberg}, {Chaplin}, {Chiappini}, {Eisenstein}, {Elsworth}, {Feuillet},
  {Fleming}, {Galbraith-Frew}, {Garc{\'\i}a}, {Garc{\'\i}a-Hern{\'a}ndez},
  {Gillespie}, {Girardi}, {Gunn}, {Hasselquist}, {Hayden}, {Hekker}, {Ivans},
  {Kinemuchi}, {Klaene}, {Mahadevan}, {Mathur}, {Mosser}, {Muna}, {Munn},
  {Nichol}, {O'Connell}, {Parejko}, {Robin}, {Rocha-Pinto}, {Schultheis},
  {Serenelli}, {Shane}, {Silva Aguirre}, {Sobeck}, {Thompson}, {Troup},
  {Weinberg}, \& {Zamora}}]{Majewski2017}
{Majewski}, S.~R., {Schiavon}, R.~P., {Frinchaboy}, P.~M., {et~al.} 2017, \aj,
  154, 94

\bibitem[{Meingast {et~al.}(2016)Meingast, Alves, Mardones, Teixeira, Lombardi,
  Gro{\ss}schedl, Ascenso, Bouy, Forbrich, Goodman, \&
  {Others}}]{Meingast2016-nb}
Meingast, S., Alves, J., Mardones, D., {et~al.} 2016, Astron. Astrophys. Suppl.
  Ser., 587, A153

\bibitem[{Murdin \& Penston(1977)}]{Murdin1977-yl}
Murdin, P. \& Penston, M.~V. 1977, Mon. Not. R. Astron. Soc., 181, 657

\bibitem[{{Price-Whelan} {et~al.}(2018){Price-Whelan}, {Sip{\H{o}}cz},
  {G{\"u}nther}, {Lim}, {Crawford}, {Conseil}, {Shupe}, {Craig}, {Dencheva},
  {Ginsburg}, {VanderPlas}, {Bradley}, {P{\'e}rez-Su{\'a}rez}, {de Val-Borro},
  {Paper Contributors}, {Aldcroft}, {Cruz}, {Robitaille}, {Tollerud},
  {Coordination Committee}, {Ardelean}, {Babej}, {Bach}, {Bachetti}, {Bakanov},
  {Bamford}, {Barentsen}, {Barmby}, {Baumbach}, {Berry}, {Biscani}, {Boquien},
  {Bostroem}, {Bouma}, {Brammer}, {Bray}, {Breytenbach}, {Buddelmeijer},
  {Burke}, {Calderone}, {Cano Rodr{\'\i}guez}, {Cara}, {Cardoso}, {Cheedella},
  {Copin}, {Corrales}, {Crichton}, {D{\textquoteright}Avella}, {Deil},
  {Depagne}, {Dietrich}, {Donath}, {Droettboom}, {Earl}, {Erben}, {Fabbro},
  {Ferreira}, {Finethy}, {Fox}, {Garrison}, {Gibbons}, {Goldstein}, {Gommers},
  {Greco}, {Greenfield}, {Groener}, {Grollier}, {Hagen}, {Hirst}, {Homeier},
  {Horton}, {Hosseinzadeh}, {Hu}, {Hunkeler}, {Ivezi{\'c}}, {Jain}, {Jenness},
  {Kanarek}, {Kendrew}, {Kern}, {Kerzendorf}, {Khvalko}, {King}, {Kirkby},
  {Kulkarni}, {Kumar}, {Lee}, {Lenz}, {Littlefair}, {Ma}, {Macleod},
  {Mastropietro}, {McCully}, {Montagnac}, {Morris}, {Mueller}, {Mumford},
  {Muna}, {Murphy}, {Nelson}, {Nguyen}, {Ninan}, {N{\"o}the}, {Ogaz}, {Oh},
  {Parejko}, {Parley}, {Pascual}, {Patil}, {Patil}, {Plunkett}, {Prochaska},
  {Rastogi}, {Reddy Janga}, {Sabater}, {Sakurikar}, {Seifert}, {Sherbert},
  {Sherwood-Taylor}, {Shih}, {Sick}, {Silbiger}, {Singanamalla}, {Singer},
  {Sladen}, {Sooley}, {Sornarajah}, {Streicher}, {Teuben}, {Thomas},
  {Tremblay}, {Turner}, {Terr{\'o}n}, {van Kerkwijk}, {de la Vega}, {Watkins},
  {Weaver}, {Whitmore}, {Woillez}, {Zabalza}, \& {Contributors}}]{astropy:2018}
{Price-Whelan}, A.~M., {Sip{\H{o}}cz}, B.~M., {G{\"u}nther}, H.~M., {et~al.}
  2018, \aj, 156, 123

\bibitem[{{Sharma} \& {Johnston}(2009)}]{enlink}
{Sharma}, S. \& {Johnston}, K.~V. 2009, \apj, 703, 1061

\bibitem[{{Strom} {et~al.}(1993){Strom}, {Strom}, \& {Merrill}}]{Strom93}
{Strom}, K.~M., {Strom}, S.~E., \& {Merrill}, K.~M. 1993, \apj, 412, 233

\bibitem[{{van der Maaten} \& Hinton(2008)}]{tsne}
{van der Maaten}, L. \& Hinton, G. 2008, Journal of Machine Learning Research,
  9, 2579

\bibitem[{{Zari} {et~al.}(2017){Zari}, {Brown}, {de Bruijne}, {Manara}, \& {de
  Zeeuw}}]{Zari2017}
{Zari}, E., {Brown}, A.~G.~A., {de Bruijne}, J., {Manara}, C.~F., \& {de
  Zeeuw}, P.~T. 2017, \aap, 608, A148

\bibitem[{{Zari} {et~al.}(2019){Zari}, {Brown}, \& {de Zeeuw}}]{Zari2019}
{Zari}, E., {Brown}, A.~G.~A., \& {de Zeeuw}, P.~T. 2019, \aap, 628, A123

\bibitem[{{Zhang} {et~al.}(2019){Zhang}, {Zhang}, {Li}, {Du}, \&
  {Habetler}}]{Zhang2019tsne}
{Zhang}, S., {Zhang}, S., {Li}, S., {Du}, L., \& {Habetler}, T.~G. 2019, arXiv
  e-prints, arXiv:1911.01024

\end{thebibliography}
	
	\newpage

	
	\begin{sidewaystable*}
	    \small
		\centering
		\begin{tabular}{rcrrlrlrrlclc}
			\hline
			SNN & Parent &
			Size & Stability &
			\begin{tabular}{@{}c@{}} $\bar{\alpha}$ \\ ($^\circ$) \end{tabular} & 
			\begin{tabular}{@{}c@{}} $\bar{\delta}$ \\ ($^\circ$) \end{tabular} &  
			\begin{tabular}{@{}c@{}} $\bar{\varpi}$ \\ (mas) \end{tabular} & 
			\begin{tabular}{@{}c@{}} $\bar{\mu_{\alpha}}$ \\ (mas\ yr$^{-1}$) \end{tabular} & 
			\begin{tabular}{@{}c@{}} $\bar{\mu_{\delta}}$ \\ (mas\ yr$^{-1}$) \end{tabular} & 
			Group name & Blaauw & Reference \\ \hline
			1  &  --  &  681  &  3393  &  81.75 (0.84)  &   2.29 (1.01)  &   2.85 (0.09)  &   1.45 (0.23)  &  -0.29 (0.33) & Briceno 1 & Ia & \cite{briceno07} \\
            2  &  --  &  149  &  3241  &  87.83 (1.01)  &  -7.92 (1.48)  &   3.34 (0.12)  &   0.19 (0.22)  &  -0.58 (0.21) & Orion Y & --- & \cite{Kounkel2018} \\
            3  &  --  &  433  &  2847  &  83.69 (1.21)  &  -1.62 (0.72)  &   2.77 (0.08)  &   1.64 (0.33)  &  -1.25 (0.23) & \textbf{OBP-near} & Ib & this work \\
            4  &  --  &  635  &  2719  &  83.92 (0.99)  &   9.77 (0.83)  &   2.47 (0.1)  &   1.26 (0.41)  &  -2.19 (0.25)  & $\lambda$ Ori & --- &  \cite{Murdin1977-yl} \\
            5  &  --  &  876  &  2196  &  83.83 (0.38)  &  -6.02 (0.64)  &   2.59 (0.09)  &    1.2 (0.21)  &   0.45 (0.24) & NGC 1980 & Ic & \cite{Alves2012} \\
            6  &  --  &  202  &  1918  &  83.21 (0.55)  &  -1.55 (0.5)  &   2.36 (0.08)  &   0.04 (0.18)  &  -0.18 (0.21)  & OBP-d & Ib & \cite{Kubiak2017-ur} \\
            7  &  --  &  141  &  1403  &  81.98 (0.6)  &   1.64 (0.58)  &   2.71 (0.08)  &  -0.58 (0.21)  &   0.75 (0.21)  & ASCC20 & Ia &  \cite{Kharchenko13} \\
            8  &  --  &  146  &  978  &  78.36 (1.01)  &  -2.43 (0.79)  &   2.61 (0.08)  &   1.39 (0.19)  &  -0.65 (0.18)  & \bf{L1616} & --- & this work \\
            9  &  --  &  123  &  947  &  84.31 (0.52)  &  -0.68 (0.52)  &   2.57 (0.08)  &  -1.11 (0.19)  &  -0.72 (0.18)  & OBP-b & Ib &  \cite{Kubiak2017-ur} \\
            10  &  --  &  78  &  739  &  86.12 (0.89)  &   7.29 (1.04)  &    2.2 (0.1)  &   -3.3 (0.19)  &  -1.72 (0.2)    & \bf{$\lambda$ Ori South} & --- & this work \\
            11  &  --  &  80  &  647  &  77.99 (1.17)  &  -8.18 (2.19)  &   3.54 (0.17)  &   1.75 (0.24)  &  -2.99 (0.22)  & \bf{Rigel} & --- & this work \\
            12  &  --  &  150  &  422  &   85.7 (0.5)  &  -8.11 (0.67)  &   2.29 (0.08)  &   0.08 (0.23)  &  -0.28 (0.22)  & L1641S & Id & \cite{Strom93} \\
            13  &  3   &  381  &  357  &  84.16 (0.89)  &  -1.53 (0.76)  &   2.77 (0.09)  &   1.78 (0.23)  &  -1.26 (0.24) & --- & --- & this work \\
            14  &  --  &  106  &  158  &  83.38 (0.74)  &   3.97 (0.93)  &   2.35 (0.08)  &  -0.94 (0.25)  &    0.6 (0.19) & \bf{ome Ori} & Ia & this work \\
            15  &  --  &  64  &  126  &  76.35 (1.03)  &  12.41 (1.2)  &    3.2 (0.15)  &    1.8 (0.29)  &  -3.55 (0.2)    & \bf{L1562} & --- & this work \\
            16  &  --  &  77  &  110  &  81.58 (0.77)  &   0.45 (0.68)  &   2.37 (0.08)  &   0.21 (0.24)  &   1.43 (0.18)  & \bf{OBP-West} & --- & this work \\
            17  &  --  &  117  &  76  &  85.58 (0.65)  &  -9.31 (0.85)  &   2.18 (0.08)  &   0.54 (0.22)  &  -1.19 (0.24)  & \bf{L1647} & --- & this work \\
            18  &  3  &  367  &  49  &   84.3 (0.8)  &  -1.63 (0.78)  &   2.76 (0.1)  &    1.8 (0.21)  &   -1.3 (0.24)     & --- & --- & this work \\
            19  &  --  &  105  &  39  &  83.83 (0.53)  &  -1.64 (0.81)  &   2.31 (0.07)  &  -1.38 (0.21)  &   0.93 (0.19)  & \textbf{OBP-far} & Ib & this work \\
            20  &  --  &  71  &  35  &  81.03 (0.39)  &  -4.29 (0.41)  &   2.61 (0.07)  &   1.93 (0.21)  &   -1.0 (0.18)   & \bf{L1634} & --- & this work\\
            21  &  --  &  53  &  27  &   88.9 (0.72)  &   0.45 (1.08)  &   2.38 (0.11)  &  -2.12 (0.17)  &   -2.1 (0.2)    & \bf{Orion B West} & --- & this work \\
            22  &  11  &  66  &  26  &  80.09 (2.25)  &  -9.94 (1.69)  &   3.53 (0.16)  &  -0.17 (0.19)  &  -1.26 (0.19)   & --- & --- & this work \\
            23  &  --  &  118  &  23  &  80.03 (1.29)  &  -4.31 (1.24)  &   3.09 (0.09)  &   0.86 (0.17)  &  -0.43 (0.15)  & Orion X & --- & \cite{Bouy2015-ce}\\
            24  &  --  &  68  &  19  &  87.39 (0.96)  &   -5.1 (0.86)  &   2.21 (0.09)  &   -2.5 (0.23)  &   0.27 (0.19)   & \bf{Orion A West} & --- & this work\\
            25  &  5  &  225  &  13  &  83.86 (0.43)  &  -6.28 (0.54)  &   2.61 (0.06)  &   1.19 (0.15)  &   0.36 (0.14)   & --- & --- & this work \\
		\end{tabular}
		\caption{Stellar groups recovered by SNN. Each row corresponds to one distinct stellar group identified by either algorithm. The names in bold highlight the newly discovered groups. The columns from left to right show for each group: 1) the SNN group index; 2) the SNN group index of the immediate group that contains, or partially contains it; 3) number of stars it contains; 4) the number of times it appears out of 7,000 SNN runs; 5) the average RA; 6) the average Dec; 7) the average parallax; 8) the average PMRA; 9) the average PMDec; 10) identifier from previous works if already known and from this work in bold if newly discovered; 11) which Blaauw structure to which it belongs; 12) reference for the identifier in 10). The numbers in parentheses from column 5) to 9) are the standard deviation of the quantities shown.}
		\label{table:finaltable}
	\end{sidewaystable*}
	
\end{document}